\documentclass[prl,twocolumn,showpacs,showkeys,amsmath,nobalancelastpage,floatfix,preprintnumbers,amssymb]{revtex4}

\usepackage[english]{babel}

\usepackage{graphicx}
\usepackage{bm}

\begin{document}

\title{Quantum circuits for general multi-qubit gates}
\author{Mikko M\"ott\"onen}
\email{mpmotton@focus.hut.fi}
\author{Juha J. Vartiainen}
\author{Ville Bergholm}
\author{Martti\ M.\ Salomaa}
\affiliation{Materials Physics Laboratory, POB 2200 (Technical Physics)\\
FIN-02015 HUT, Helsinki University of Technology, Finland}
\date{\today}

\begin{abstract}
We consider the minimal elementary gate sequence which is needed to implement a general quantum gate acting on $n$ qubits --- a unitary transformation with $4^n$ degrees of freedom. For synthesizing the gate sequence, a method based on the so-called cosine-sine matrix decomposition is presented. The result is optimal in the number of elementary one-qubit gates, $4^n$, and scales more favourably than the previously reported decompositions requiring $4^n - 2^{n+1}$ CNOT gates.
\end{abstract}


\maketitle

\parindent 0mm
\parskip 5mm

The foundation of quantum computation~\cite{nc} involves the encoding of computational tasks into the temporal evolution of a quantum system. Thereby a register of $n$~qubits, identical two-state quantum systems, is employed. Quantum algorithms can be described by unitary transformations and projective measurements acting on the $2^n$-dimensional state vector of the register. In this context, unitary transformations are also called quantum gates. The recently discovered quantum algorithms \cite{Abrams,Jaksch,Paz} embody arbitrary unitary transformations and hence call for techniques to efficiently implement a general $n$-qubit gate. The complexity of an implementation is measured in terms of the number of elementary gates required~\cite{barenco}. Achieving gate arrays of lower complexity is crucial not only because it generally results in shorter execution times, but it may also introduce less errors.

Any finite-dimensional unitary transformation can be represented as a unitary matrix and hence any $n$-qubit gate corresponds to a certain $2^n \times 2^n$ unitary matrix,~$U$. Therefore, the powerful methods of matrix computation~\cite{golub} can be utilized to produce quantum gate decompositions. However, only decompositions yielding matrices which correspond to gate sequences of low complexity are interesting. We choose the library of elementary gates to consist of the controlled-NOT (CNOT) gate, the one-qubit rotations about the $y$ and $z$~axes, and a phase gate adjusting the unobservable global phase.  Since the cost of physically realizing a CNOT gate may exceed that of a one-qubit gate, we count the numbers of these gates separately.

A general unitary $2^n \times 2^n$ matrix $U$ has $4^n$~real degrees of freedom. Since each elementary one-qubit gate carries one degree of freedom, at least $4^n$ such gates are needed to implement $U$. The current theoretical lower bound for the number of CNOT gates needed in realizing an arbitrary $n$-qubit gate, $\lceil \frac{1}{4}(4^n-3n-1) \rceil$, is given in Ref.~\cite{shende}. However, no circuit construction yielding these numbers of CNOT or elementary one-qubit gates has been presented in the literature. The conventional approach~\cite{barenco} to implementing general multi-qubit gates makes use of the QR decomposition~\cite{golub} for unitary matrices, yielding an array of $O(n^34^n)$ elementary gates. Heretofore, the most efficient implementation based on the QR decomposition, for asymptotically large~$n$, requires approximately $8.7 \cdot 4^n$ CNOT gates~\cite{PRL}. In addition, the synthesis of optimal quantum circuits for certain special classes of gates has been intensively studied. The implementation of a general two-qubit gate~\cite{shende,vatan2,whaley_minimum,vidal} is found to require 3~CNOTs and 16~elementary one-qubit gates. For a three-qubit gate, the current minimal implementation using 40~CNOTs and 98~elementary one-qubit gates~\cite{vatan3} is based on the Khaneja-Glaser decomposition (KGD)~\cite{khaneja}. Furthermore, an implementation of an arbitrary diagonal unitary matrix involving $2^n-2$ CNOTs and $2^n$ elementary one-qubit gates is known~\cite{bullock:2004}.

In this Letter, we present an efficient implementation of a general unitary transformation $U$ by recursively utilizing the cosine-sine decomposition (CSD)~\cite{Paige}. In the context of quantum computation, the CSD has first been considered in Ref.~\cite{tucci}, and its relation to the KGD has recently been discussed in~\cite{bullock2}. We decompose~$U$ into a product of matrices, each of which is identified with a new type of gate which we call a uniformly controlled rotation. To implement these gates, we present an efficient elementary gate sequence which is related to the gates recently explored in Ref.~\cite{bullock:2004} as a part of the implementation of a diagonal quantum computer.

\begin{figure}
\includegraphics[width=0.45\textwidth]{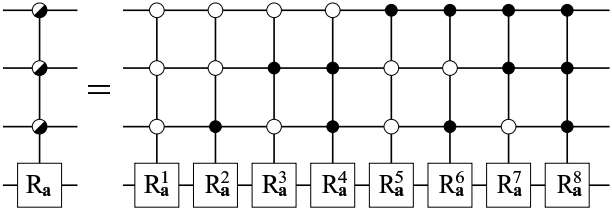}
\caption{\label{fig:F} Definition of the uniformly controlled rotation $F^3_4(R_{\bf a})$. Here ${\bf a}$ is a
three-dimensional vector fixing the rotation axis of the matrices $R_{\bf a}^j=R_{\bf a}(\alpha_j)$.}
\end{figure}

Let $F^k_m(R_{\bf a})$ denote a uniformly controlled rotation. It consist of $k$-fold controlled rotations of qubit~$m$ about the three-dimensional vector ${\bf a}$, one rotation for each of the $2^k$ different classical values of the control qubits. The index $m$ may acquire the values $1,2,\dots,n$  and $k$ the values $1,2,\dots,n-1$. An example of $F^k_m(R_{\bf a})$, where $m=4$ and $k=3$ is shown in Fig.~\ref{fig:F}. The relative order of the controlled rotations is irrelevant; the gates commute. For instance, the uniformly controlled rotation $F^k_{k+1}(R_{\bf a})$ has the matrix representation
\begin{equation}
F^k_{k+1}(R_{\bf a}) =
\begin{pmatrix}
R_{\bf a}(\alpha_1)\\
& \ddots &\\
& & R_{\bf a}(\alpha_{2^k})
\end{pmatrix},
\end{equation}
where the angles $\alpha_1, \alpha_2, \ldots, \alpha_{2^k}$ may be freely chosen and the rotation matrix $R_{\bf a}(\phi)$ is given by 
\begin{equation}\label{eq:ra} 
R_{{\bf a}}(\phi) = e^{i  {\bf a} \cdot \bm{\sigma}\phi/2} =
I \cos \frac{\phi}{2} + i \left( {\bf a} \cdot \bm{\sigma}
\right)\sin \frac{\phi}{2}.
\end{equation}
Above $I$~is the unit matrix and the product ${\bf a} \cdot\bm{\sigma}=a_x\sigma_x+a_y\sigma_y+a_z\sigma_z$ involves the Pauli matrices $\sigma_x$, $\sigma_y$, and $\sigma_z$~\cite{nc}. In general, $F^k_m(R_{\bf a})$ is a product of $2^k$ two-level matrices.

\begin{figure}
\includegraphics[width=0.45\textwidth]{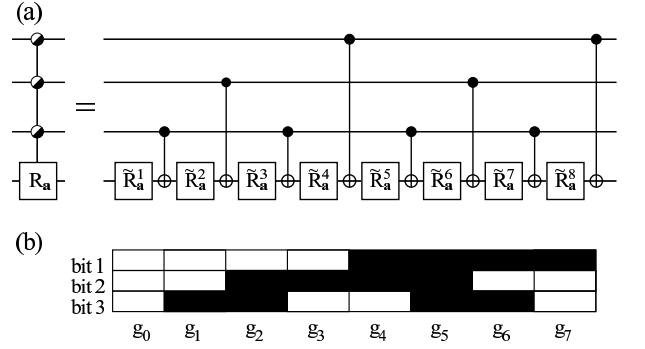}
\caption{\label{fig:piiri}
(a) Quantum circuit realizing the gate $F^3_4(R_{{\bf a}})$, where ${\bf a}$ is perpendicular to the $x$-axis. Here we have used a notation $\tilde{R}_{\bf a}^j=R_{\bf a}(\theta_j)$. (b) Binary reflected 3-bit Gray code used to define the positions of the control nodes. The black and white rectangles denote bit values one and zero, respectively.}
\end{figure}

We propose an implementation of $F^k_m(R_{\bf a})$ with $a_x=0$ using an alternating sequence of $2^k$~CNOTs and $2^k$~one-qubit rotations $R_{{\bf a}}(\theta_i)$ acting on the qubit~$m$. The position of the control node in the $l^{\rm th}$ CNOT gate is set to match the position where the $l^{\rm th}$ and $(l+1)^{\rm th}$ bit strings $g_{l-1}$ and $g_{l}$ of the binary reflected Gray code~\cite{savage:1997} differ. In binary Gray codes, the adjacent bit strings differ by definition only in a single bit, and hence the position is well defined. As an example, the quantum circuit for the gate $F^3_4(R_{{\bf a}})$ is shown in Fig.~\ref{fig:piiri}(a) while Fig.~\ref{fig:piiri}(b) illustrates the correspondence of the Gray code to the positions of the control nodes in the CNOT gates.

In the proposed construction, each of the control qubits regulates an even number of NOT gates, since in a cyclic Gray code each bit is flipped an even number of times. On the other hand, Eq.~(\ref{eq:ra}) yields
\begin{equation}
a_x = 0 \implies \sigma_x R_{{\bf a}}(\theta) \sigma_x =
R_{{\bf a}}(-\theta).
\label{eq:konjugaatio}
\end{equation}
Hence, for any of the standard basis vectors acting as an input all the induced NOT gates annihilate each other and negate some of the angles $\{\theta_i\}$. Furthermore, subsequent rotations about any single axis ${\bf a}$ are additive, i.e., $R_{{\bf a}}(\phi) R_{{\bf a}}(\omega) = R_{{\bf a}}(\phi + \omega)$ for arbitrary angles $\phi$ and $\omega$. Thus, the construction yields a rotation of the qubit $m$ about the axis ${\bf a}$ through an angle which is a linear combination of the angles $\{\theta_i\}$. Consequently, the proposed quantum circuit is equivalent to the gate $F^k_m(R_{{\bf a}})$ provided that the angles $\{\theta_i\}$ are a solution of the linear system of equations
\begin{equation}
M^{k}
\begin{pmatrix}
\theta_1 \\
\vdots \\
\theta_{2^k}
\end{pmatrix}
=
\begin{pmatrix}
\alpha_1 \\
\vdots \\
\alpha_{2^k}
\end{pmatrix},
\label{eq:linear}
\end{equation}
where the matrix elements $M^{k}_{ij}$ can be determined using Eq.~(\ref{eq:konjugaatio}). The rotation angle~$\theta_j$ is negated, provided that the control nodes attached to the $l^{\text{th}}$ qubit are active and the $l^{\text{th}}$~bit of $g_{j-1}$ has the value one. The negations must be applied for each control qubit independently, which results in
\begin{equation}
M^{k}_{ij} = (-1)^{b_{i-1} \cdot g_{j-1}},
\end{equation}
where $b_i$ is the standard binary code representation of the integer $i$ and the dot in the exponent denotes the bitwise inner product of the binary vectors.

The matrix $M^k$ bears a strong resemblance to the $k$-bit Walsh-Hadamard matrix $H^{k}_{ij} = 2^{-k/2} (-1)^{b_{i-1} \cdot b_{j-1}}$, which is by construction orthogonal. Since a Gray code is a permutation of the standard binary code, $2^{-k/2} M^{k}$ is a column-permuted version of $H^{k}$ and thus also orthogonal. Consequently, we obtain the inverse matrix $(M^{k})^{-1}=2^{-k}(M^{k})^{T}$ and the determination of $\{\theta_i\}$ for any desired angles $\{\alpha_i\}$ is immediate. Thus any uniformly controlled rotation $F^k_m(R_{{\bf a}})$ with $a_x=0$ and $k \ge 1$ can be realized using $2^k$~CNOT gates and $2^k$~one-qubit rotations $R_{{\bf a}}(\theta_i)$. We note that although we chose to use the binary reflected Gray code to determine the positions of the control nodes in the CNOT gates, any cyclic $k$-bit binary Gray code will also qualify. Furthermore, $F^k_m(R_{{\bf a}})$ can also be achieved by a horizontally mirrored version of the quantum circuit presented.

The CSD of a unitary $2^n \times 2^n$ matrix may be expressed as~\cite{Paige}
\begin{equation}\label{eq:cs}
U =
\underbrace{\begin{pmatrix}
u^1_{11} & 0\\
0 & u^1_{12}
\end{pmatrix}}_{U^1_1}
\underbrace{
\begin{pmatrix}
c^1_{11} & s^1_{11}\\
-s^1_{11} & c^1_{11}
\end{pmatrix}
}_{A^1_1}
\underbrace{\begin{pmatrix}
u^1_{21} & 0\\
0& u^1_{22}
\end{pmatrix}}_{\tilde{U}^1_2},
\end{equation}
where the exact form of the submatrices is given below.
The decomposition may be applied recursively to the submatrices of $U^i_j$, until a $2\times 2$ block-diagonal form is encountered. In our indexing scheme, the upper index denotes the level of recursion, whereas the lower index denotes the position of the matrix within the resulting matrix product. We note that CSD is not unique, and one should take the possible internal symmetries of the matrix $U$ into account to obtain the simplest achievable form for the matrices~$U^i_{j}$.

In the decomposition, $u^i_{jk}$ ($k = 1, \ldots, 2^i$) are unitary $2^{n-i} \times 2^{n-i}$ matrices and the real diagonal matrices $c^i_{jk}$ and $s^i_{jk}$ ($k = 1, \ldots, 2^{i-1}$) are of the form $c^i_{jk} = \text{diag}_{l}(\cos \theta_{l})$ and $s^i_{jk} = \text{diag}_{l}(\sin \theta_{l})$ ($l = 1, \ldots, 2^{n-i}$).
For a general $i = 1, \,\ldots, n-1$, the matrices $U^i_j$ and $A^i_j$ assume the forms
\begin{equation}\label{eq:ycs}
U^i_{j}={\rm diag}_k(u_{jk}^i); \quad (k = 1, \ldots, 2^i),
\end{equation}
and
\begin{equation}
A^i_{j}={\rm diag}_{k}\left[
\begin{pmatrix}
c_{jk}^{i} & s_{jk}^{i} \\
-s_{jk}^{i} & c_{jk}^{i}
\end{pmatrix}
\right]; \quad (k = 1, \ldots, 2^{i-1}),
\end{equation}
where the Eq.~(\ref{eq:ycs}) applies also for $\tilde{U}^i_{j}$.
For the $i^{\text{th}}$ level of the recursion we obtain
\begin{equation}
\label{eq:ics}
U^{i-1}_{j} = U^{i}_{2j-1} A^i_{\zeta(i,j)} \tilde{U}^{i}_{2j},
\end{equation}
where the indexing function $\zeta(i,j) = 2^{n-i-1}(2j - 1)$ has been introduced to make the result of the recursion more feasible. The matrix $A_1^1$ is also referred to as $A_{\zeta(1,1)}^1$. As compared with the original matrix $U^{i-1}_j$, the above decomposition contains $2^{n-1}$ additional degrees of freedom. To specify them explicitly, we define unitary diagonal matrices
\begin{equation}
P^i_j = \text{diag}_k(p^i_{j,\lceil k/2 \rceil}),
\end{equation}
where $j,k = 1,\ldots,2^i$ and the diagonal matrix $p^i_{jk} = \text{diag}_l(e^{i \alpha_l})$, where $l=1,\ldots,2^{n-i}$.
The angles $\{\alpha_l\}$ may be chosen arbitrarily for each $p^i_{jk}$ and, as shown below, we can use them to reduce the total number of gates needed in the final decomposition.
We insert $I = P^i_{\zeta(i,j)} (P^i_{\zeta(i,j)})^{\dagger}$ into Eq.~(\ref{eq:ics}), next to $A^i_{\zeta(i,j)}$ with which $P^i_{\zeta(i,j)}$ commutes, and obtain
\begin{equation}
\label{eq:xcs}
U^{i-1}_{j} = U^{i}_{2j-1} P^i_{\zeta(i,j)} A^i_{\zeta(i,j)} U^{i}_{2j},
\end{equation}
where the matrix $(P^i_{\zeta(i,j)})^{\dagger}$ is absorbed into the definition of $U^i_{2j} = (P^i_{\zeta(i,j)})^{\dagger} \tilde{U}^i_{2j}$, and $P^i_{\zeta(i,j)}$ is kept intact.

Finally, the decomposition leads to the result
\begin{equation}\label{eq:dec}
U = \left( \prod_{j=1}^{2^{n-1}-1} \underbrace{U^{n-1}_j P^{\gamma(j)}_{j}}_{B_j}
A^{\gamma(j)}_{j} \right) \underbrace{U^{n-1}_{2^{n-1}}}_{B_{2^{n-1}}},
\end{equation}
where the function $\gamma(j) + 1$ indicates the position of the least significant non-zero bit in the $n$-bit binary presentation of the number~$j$. The matrices $P^{\gamma(j)}_{j}$ are determined by the preceding matrix $U^{n-1}_j$. This fixes the order in which the recursion must be applied, since the absorbed matrices $(P^{\gamma(j)}_{j})^{\dagger}$ affect consequent decompositions. Thus, the recursion in Eq.~(\ref{eq:xcs}) is first applied to the matrix $U^i_j$ with the largest upper index and, upper indices being equal, the smallest lower index subject to the stopping criterion $i = n-1$. 

\begin{figure*}
\includegraphics[width=0.92\textwidth]{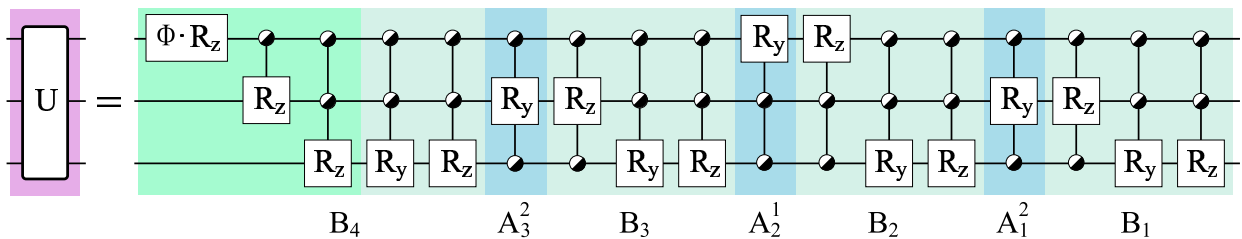}
\caption{Quantum circuit for a three-qubit gate obtained using the cosine-sine decomposition. The sequences of gates $B_j$ correspond to the $2\times 2$-block diagonal matrices and the gates $A^i_j$ to the cosine-sine matrices. The leftmost gate sequence corresponds to the diagonal quantum computer of Ref.~\cite{bullock:2004}.}
\label{fig:hajo}
\end{figure*}

We find that each of the matrices $A^i_j$ in Eq.~(\ref{eq:dec}) corresponds to a gate $F^{n-1}_i(R_y)$. Furthermore, the $2 \times 2$ block-diagonal matrices~$B_j$ may, with a suitable choice of $P^{\gamma(j)}_{j}$, be expressed as
\begin{equation}
B_j = F^{n-1}_n(R_z) F^{n-1}_n(R_y) F^{n-1}_{\gamma(j)}(R_z),
\end{equation}
and combined with the subsequent $A^{\gamma(j)}_{j}$ into a $B\!A$ section:
\begin{equation}
(B\!A)_j =
F^{n-1}_n(R_z) F^{n-1}_n(R_y) F^{n-1}_{\gamma(j)}(R_z) F^{n-1}_{\gamma(j)}(R_y).
\end{equation}
The final matrix $B_{2^{n-1}}$, for which we have no extra degrees of freedom left, must be implemented as
\begin{align}
B_{2^{n-1}} =\, & F^{n-1}_n(R_z) F^{n-1}_n(R_y)  \nonumber \\ 
&\times F^{n-1}_n(R_z)F^{n-2}_{n-1}(R_z) \cdots F^{0}_1(R_z) \Phi,
\end{align}
where $\Phi$ is an elementary phase gate which serves to fix the unobservable global phase. To illustrate the method, the complete decomposition of a general three-qubit gate is shown in Fig.~\ref{fig:hajo}.

Each of the $B\!A$~sections consists of two uniformly controlled $z$~rotations and two uniformly controlled $y$~rotations. By mirroring the circuits of the $y$~rotations, we may cancel four CNOT gates in each section. Hence the cost of each of the $2^{n-1} - 1$ sections is $2^{n+1}$~elementary one-qubit rotations and $2^{n+1} - 4$ CNOTs. The final $B$~matrix decomposes into uniformly controlled $z$ and $y$~rotations followed by a cascade of uniformly controlled $z$~rotations which fixes the phases. This cascade corresponds to the diagonal quantum computer of Ref.~\cite{bullock:2004}. Applying the mirroring trick, two more CNOT gates are cancelled between the $z$ and $y$~rotations. The cost of the last $B$~section is $2^{n+1}$ elementary one-qubit gates and $2^{n+1} - 4$ CNOTs. Finally, we arrive at the total complexity of the decomposition: $4^n - 2^{n+1}$ CNOT gates and $4^n$~elementary one-qubit gates.

In conclusion, the proposed decomposition of a general multi-qubit gate, based on the CSD and uniformly controlled rotations, provides a quantum circuit that contains the minimal number of elementary one-qubit gates and on the order of four times the minimal number of CNOT gates. Compared with the minimal decomposition of a two-qubit gate~\cite{shende,vatan2,whaley_minimum,vidal} the CSD method requires 5 extra CNOT gates. For a three-qubit gate the CSD requires 48 CNOT gates and 64~elementary one-qubit gates, as opposed to the circuit of 40 CNOTs and 98~elementary one-qubit gates obtained using the KGD in Ref.~\cite{vatan3}. For four-qubit gates the CSD provides a quantum circuit of 256~elementary one-qubit gates and 224~CNOTs, which is the shortest elementary gate array known to implement such a gate. Thus, for a general $n$-qubit gate, where $n \geq 4$, the method presented provides the most efficient quantum circuit known to implement the gate.

To further improve the implementation of a particular quantum gate one may optimize the synthesized quantum circuit. The possible methods for optimization include finding the most efficient CSD factorizations, varying the Gray codes, mirroring the gate arrays of the uniformly controlled rotations and possibly combining the uniformly controlled $y$ and $z$ rotations into general uniformly controlled gates. Certain quantum gates that are likely to be useful in quantum computation comprise internal symmetries and can thus be implemented using only a polynomial number of elementary gates. For example, $O(n^2)$ gates are needed to implement a quantum Fourier transformation (QFT) of $n$~qubits \cite{nc}. Although the method presented apparently requires $O(4^n)$ elementary gates, it is still possible that using proper optimizations the gate array will appreciably simplify and the result will resemble that of the polynomial decompositions. 

\begin{acknowledgments}
This research is supported by the Academy of Finland through the project ``Quantum Computation'' (No. 206457). MM and JJV thank the Foundation of Technology (Finland), JJV the Nokia Foundation, MM and VB the Finnish Cultural Foundation, and MMS the Japan Society for the Promotion of Science for financial support. S.~M.~M.~Virtanen is acknowledged for stimulating discussions.
\end{acknowledgments}

\bibliography{csd}

\end{document}